\documentclass{PoS}

\newcommand{\beq}{\begin{equation}}
\newcommand{\eeq}{\end{equation}}
\newcommand{\bea}{\begin{eqnarray}}
\newcommand{\eea}{\end{eqnarray}}

\newcommand{\nn}{\nonumber}
\newcommand{\e}{\rm e\,}
\newcommand{\tr}{\hbox{tr}}

\newcommand{\I}{\hbox{Im}}
\newcommand{\R}{\hbox{Re}}

\title{Systematic approximation for QCD at non-zero density}

\ShortTitle{Systematic approximation for QCD at non-zero density}

\author{\speaker{Ion-Olimpiu Stamatescu}
\\
        Institut f\"ur Theoretische Physik, Universit\"at Heidelberg,
	Heidelberg, Germany\\
        E-mail: \email{stamates@thphys.uni-heidelberg.de}}
\author{Gert Aarts\\
        Department of Physics, College of Science, Swansea University,
	 Swansea, UK\\
        E-mail: \email{g.aarts@swansea.ac.uk}}
\author{Benjamin J\"ager\\
        Department of Physics, College of Science, Swansea University, 
	Swansea, UK\\
        E-mail: \email{B.Jaeger@swansea.ac.uk}}
\author{Erhard Seiler\\
        Max-Planck-Institut f\"ur Physik (Werner-Heisenberg-Institut), 
M\"unchen, Germany\\
        E-mail: \email{ehs@mppmu.mpg.de}}
\author{Denes Sexty\\
        Department of Physics, Bergische Universit\"at Wuppertal, 
	Wuppertal, Germany\\
        E-mail: \email{d.sexty@thphys.uni-heidelberg.de}}

\abstract{We use the heavy dense formulation of QCD (HD-QCD) as the
basis for an analytic expansion as systematic
approximation
to QCD at non-zero density,  keeping the 
full Yang-Mills action.  
We analyse the structure of the baryonic density and other 
quantities and
 present data from the complex Langevin
equation (CLE) and reweighting (RW) calculations for 2 flavours of 
Wilson fermions.}

\FullConference{The 32nd International Symposium on Lattice 
Field Theory,\\
		23-28 June, 2014\\
		Columbia University New York, NY}

\begin{document}

\section{HD-QCD as basis for a systematic analytic approximation to QCD}

The QCD grand canonical partition function for $n_{f} =1$ flavour 
of Wilson fermions is:
{\bea
Z &=&  \int DU\,\rho, \quad \rho = \e^{-S} =  \e^{-S_{YM}} \det M 
\label{e.zqcd}\\
{\rm M} &=& 1 - 2\,\kappa\sum_{i=1}^3 \left( \Gamma_{+i} U_{x,i}T_{i} 
+ \Gamma_{-i} U_{x,i}^{\dagger} T_{-i} \right) 
-2\,\kappa \gamma \left( \e^\mu \Gamma_{+4} U_{x,4}T_{4} +
\e^{-\mu}\Gamma_{-4} U_{x,4}^{\dagger} T_{-4} \right)
\label{eqMQCD} 
\eea}
with $\Gamma_{\pm \mu} = \frac{1}{2}(1 \pm 
\gamma_\mu)$,
$T$: lattice translations, $\kappa$: hopping parameter,
$\mu$ chemical potential,
$\gamma$:  anisotropy parameter. The temperature is introduced as 
 $\  aT=\frac{\gamma}{N_\tau}\ $ ($\gamma=1$ below).

HD-QCD  \cite{bender,blum,karsch} relies on 
the double limit
\bea
\kappa \rightarrow 0, \  \mu \rightarrow \infty , \ \ \zeta = 2 \kappa \,
\e^{\mu} \, : fixed\, . \label{e.hadmlim}
\eea
In the 0-th order in $\kappa$ (LO) only the
 Polyakov loops $P$ survive in the loop expansion
  for the fermionic 
deteminant which becomes a product of local terms
\bea
&&\det {\rm M}^0(\mu) 
= \left(1+C^3\right)^2 \left(1+{C'}^3\right)^2
\prod_{\vec x} \left(1 + a\, P_{\vec x} + b\, P_{\vec x}' \right)^2
\left(1 + {\tilde a}\, P_{\vec x}' + {\tilde b}\, P_{\vec x} \right)^2 
\ \label{detm0}\\
&& a= 3 C/(1+C^3), \, b= 3 C^2/(1+C^3),\,
{\tilde a}= 3 C'/(1+{C'}^3), \, {\tilde b}= 3 {C'}^2/(1+{C'}^3)\\
&&C=\left(2 \kappa \e^{\mu}\right)^{N_{\tau}}, \ \
C'=\left(2 \kappa \e^{-\mu}\right)^{N_{\tau}}, 
\quad P=\frac{1}{3}\tr\prod_t U_t,\ \ P'=\frac{1}{3}\tr\prod_t U_t^{-1}
\eea 
where we
also introduced the non-dominant factors from the 
inverse Polyakov loops $P'$ to preserve
 the $\det {\rm M}(\mu) = [\det {\rm M}(-\mu)]^*$ symmetry.
The LO 
describes gluonic interactions in a 
  background of static charges. Note that $a, b$ have  maxima of $2^{2/3}$
 at $ \mu = \mp \frac{1}{3 N_{\tau}} - \ln(2 \kappa)$.

\begin{figure}[h]
\begin{center}
\includegraphics[width=0.5\columnwidth]{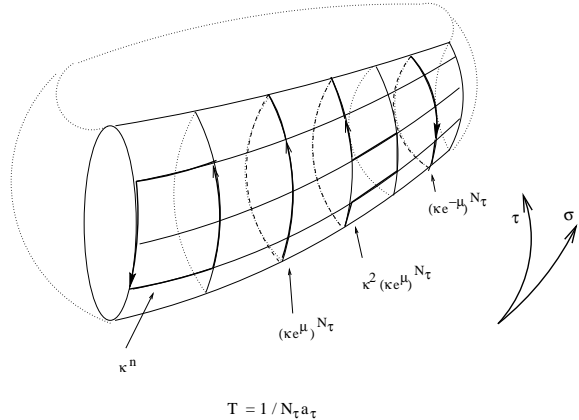}  
\caption{HD - QCD: LO and NLO.
}
\label{HDMtorus}
\end{center}
\end{figure}

   We can go beyond the 
  static limit (LO) with successive N$^q$LO
  to approximate full QCD using the analytic hopping expansion.  
  
 For $q=1$ the NLO can be defined in the loop 
 expansion using 
decorated 
Polyakov loops, for explicit formulae see 
\cite{Aarts:2002,dfss}.
 The determinant still factorizes, but the quarks have some
mobility. 
Higher N$^q$LO become 
increasingly cumbersome in the loop expansion, 
however, because of the combinatorics. 
The
effective expansion parameter is  $N_{\tau} \kappa$.

 Alternatively we can 
 expand to higher orders "algebraically"  \cite{asss}, see also 
 \cite{Langelage:2014vpa}. 
In the following we
 use two expansions, called the $\kappa$- and the $\kappa_s$-expansion.
 To define them one separates temporal and spatial hoppings 
 in the fermionic matrix M:
 \bea
&&{\rm M} \ =\ 1 -\kappa Q \ =\ 1 - R - \kappa_s S ,
\quad R = \kappa Q_t, \quad S=Q_s\\
&&\kappa{\rm-expansion}: \quad 
\det {\rm M} =  \exp \left\{-\sum_n\frac{\kappa^n}{n} Q^n\right\}\\
 &&\kappa_s{\rm-expansion}: \quad 
\det {\rm M} = \det {\rm M}^0\exp \left\{-\sum_n\frac{\kappa_s^n}{n} 
\left(\frac{S}{1-R}\right)^n
\right\}
\eea
The drifts for CLE in N$^q$LO can be systematically
 derived and used for simulations with the
 full YM action to any desired order $q$ \cite{asss}. 
 In the calculations one keeps  $\kappa_s=\kappa$.
   
  HD-QCD has been used as a model in LO and  NLO to study the
  phase diagram and other properties with  the full YM action 
  \cite{Aarts:2002,dfss,sss} and in strong coupling also 
  to higher orders \cite{phip}.

\section{Simulation method}

\subsection{Complex Langevin Simulation}

The Complex Langevin Equation (CLE) has the potential
to simulate lattice models with a complex action and for which 
 usual importance sampling 
 fails.  To  develop it to a reliable method
 is  both rewarding and tough. 

The LE is a stochastic process in which the updating of the
variables is achieved by addition of a drift term (or "force") and 
a suitably normalized random noise.
For a complex action the drift is also complex and this
automatically provides an imaginary part for the field.
This implies setting up the problem in the complexification 
of the original manifold  $R^n \longrightarrow
C^n$ or $SU(n) \longrightarrow SL(n,C)$.
The CLE then amounts to  two related, real LE with independent 
noise terms - 
here in compact form for just one variable $z = x + i\, y$ and with
$K= - \partial_z S(z)$:
 \bea
{\delta z(t)} &=& K(z)\,\delta t + \sqrt{N_R}\,
\eta_R+ {\rm i}\,\sqrt{N_I}\,\eta_I \nn \\
  \langle \eta_R\rangle &=& \langle\eta_I\rangle =0\,,\  
  \langle \eta_R \eta_I\rangle =0 
 \, , \quad\langle \eta_R^2\rangle =\langle \eta_I^2\rangle = 
 2\,\delta t\,,\ 
\, \ N_R - N_I =1\nn
\eea
In the simulations we shall take $N_I=0$. The probability distribution $P(x,y;t) $ realized in the process 
evolves according to a real Fokker-Planck equation:
\bea
\partial_t  P(x,y,t) =L^T P(x,y,t)\, ,  \quad L =
(N_R\partial_x + {\R} K(z))\partial_x + ( N_I\partial_x + 
{\I} K(z))\partial_y
\eea
One can also define a complex distribution $\rho(x,t)$  
\bea
\partial_t \rho(x,t) = L^T_0 \rho(x,t)\,, \quad  L_0 = 
(\partial_x + K(x))\partial_x   \nn
\eea
with the asymptotic solution $\rho(x)\simeq \exp(-S(x))$ and 
formally prove for the observables $O(z)$
\bea \quad
\int {O}(x+iy) P(x,y;t) dxdy = 
\int  {O}(x) \rho(x;t) dx . \nn
\eea
As for any method the proof of 
convergence depends on some 
conditions, for CLE these include:
 
 - rapid decay of $P(x,y,t)$ in $y$ \cite{trust} and 

 - holomorphy of the drift and of the observables. 

There are, moreover, a number of numerical problems, such as runaways. 
Many of them are due to the amplification of numerical imprecisons by 
unstable modes in the drift dynamics. For further discussion see, e.g.,
\cite{stab} and the references therein.

To deal with these problems one can use the freedom in 
defining the process
for a given action and the symmetries of the latter. For 
gauge theories
we set up a method ("gauge cooling") to obtain a narrow 
$y-$distribution. 
Together with using adaptive step size this also eliminates 
runaways and 
divergences triggered by numerical imprecisions. For a review see
\cite{review}.

Zeroes in the original measure  $\rho(z)$ 
lead to
a meromorphic drift. The poles in the drift can cause 
wrong convergence of the process, as shown 
in nontrivial, soluble models \cite{kim}. This problem 
is presently under study. In the cases of physical 
interest the effects due to poles do not
appear quantitatively relevant, however a systematic understanding
is still missing.

For QCD at  $\mu \neq 0$ the method has been defined in 
\cite{aarsta} and applied, 
e.g., in \cite{sss,phip,den13,ben14}.

\subsection{Reweighting}

For completeness we shall briefly describe the reweighting method (RW) 
also applied in this study. It has been used in the HD-QCD context before
\cite{dfss} but without the second factor in (\ref{detm0}) (which is
$\simeq 1$  at  large $\mu$ but is relevant if we want
to connect to the small $\mu$ region). 

We split the Boltzman factor in (\ref{e.zqcd}) and calculate
 the expectation values by reweighting
\bea
&&\rho \equiv \e^{- S_{YM}} \det M = H \, w, \quad H >0 ,\quad \quad
\langle O \rangle_{\rho} = \frac{\langle O \, w\rangle_{H}}
{\langle w\rangle_{H}} \label{e.rw}
\\
&& {\rm H} = \e^{-S_{YM} + C\,\tr P +C'\,\tr P'}, 
\quad w =  \e^{-C\,\tr P -C'\,\tr P'}  \det M
\eea
by taking into the updating factor H part of the LO 
determinant eq. (\ref{detm0}). This H allows a fast updating in producing 
the ensemble, e.g. in maximal gauge, at least to LO and NLO, since 
the additional terms can just be 
added to the staples and  used in heat bath updating.

\section{Tests and preliminary results}

\subsection{Simulations}

As  previously discussed \cite{sss} for the present CLE simulations 
 a reliability lower threshold at $\beta
 \simeq 5.7$ appears to hold. Large $\beta$ are unproblematic, large 
 $\mu$ and large $N_{\tau}$ are under study.
 
 With RW we can go to  lower $\beta$ but the signal/noise ratio strongly 
 decreases.
  Moreover RW cannot reach large chemical potential
 $\mu \simeq 1$ where the sign problem becomes acute. 
 
 In the following we work at $\beta =5.8$ and $5.9$ using 2 degenerate
 flavours of 
 Wilson fermions,  $n_f=2$. We use $\kappa=0.12$
for which the $\kappa$ expansion is expected to converge. It turns 
out that for higher
values for $\kappa$ HD-QCD may not lead to a well controlled approximation
for QCD. We use
$N_{\tau}  =8,10,12,16$ except for the comparison with full QCD where we 
use a lattice of $4^4$. The full QCD data are obtained 
by CLE \cite{den13},
the HD-QCD data by CLE and RW.


 \subsection{Approaching full QCD in the N$^q$LO series of HD-QCD}

 Since  even higher orders N$^q$LO of HD-QCD are significantly easier 
  to simulate than the full theory we want to estimate 
 how well we approximate the latter in this way. 
 
\begin{figure}[h]
\begin{center}
\includegraphics[width=0.49\columnwidth]{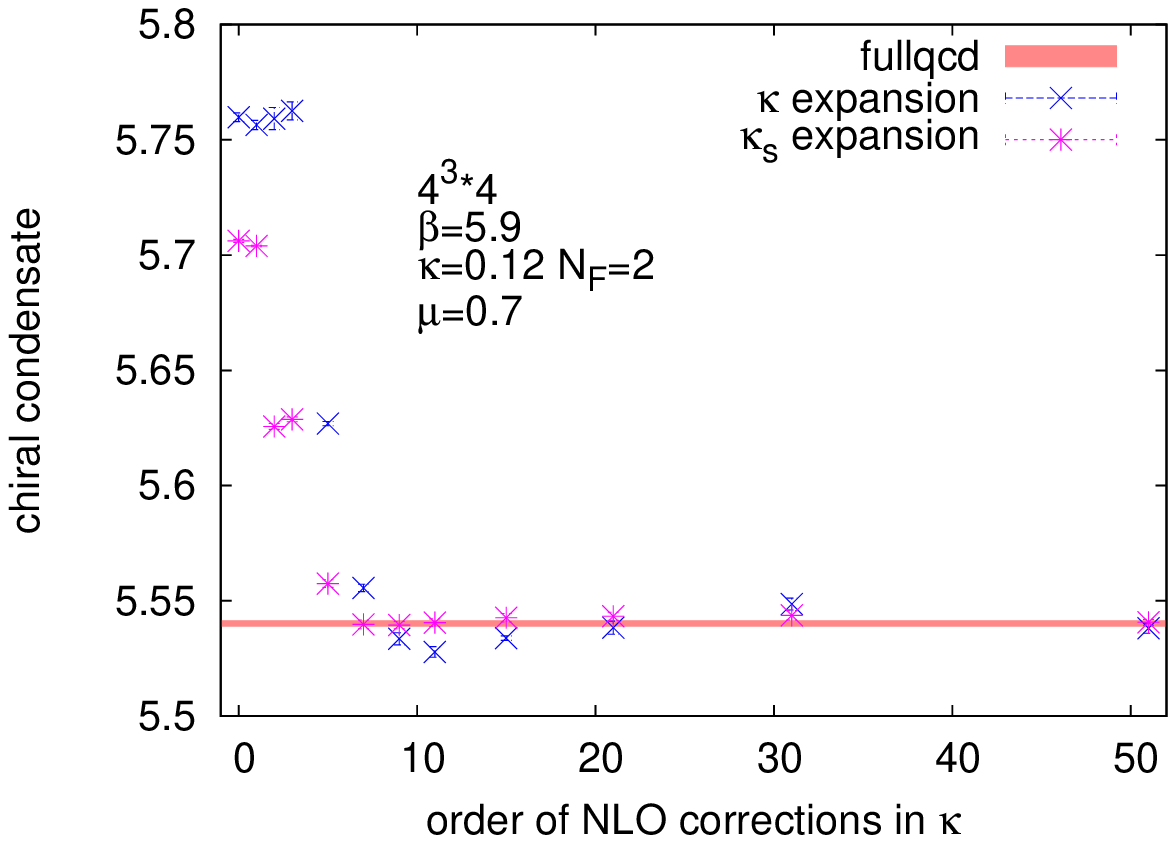}  
\includegraphics[width=0.49\columnwidth]{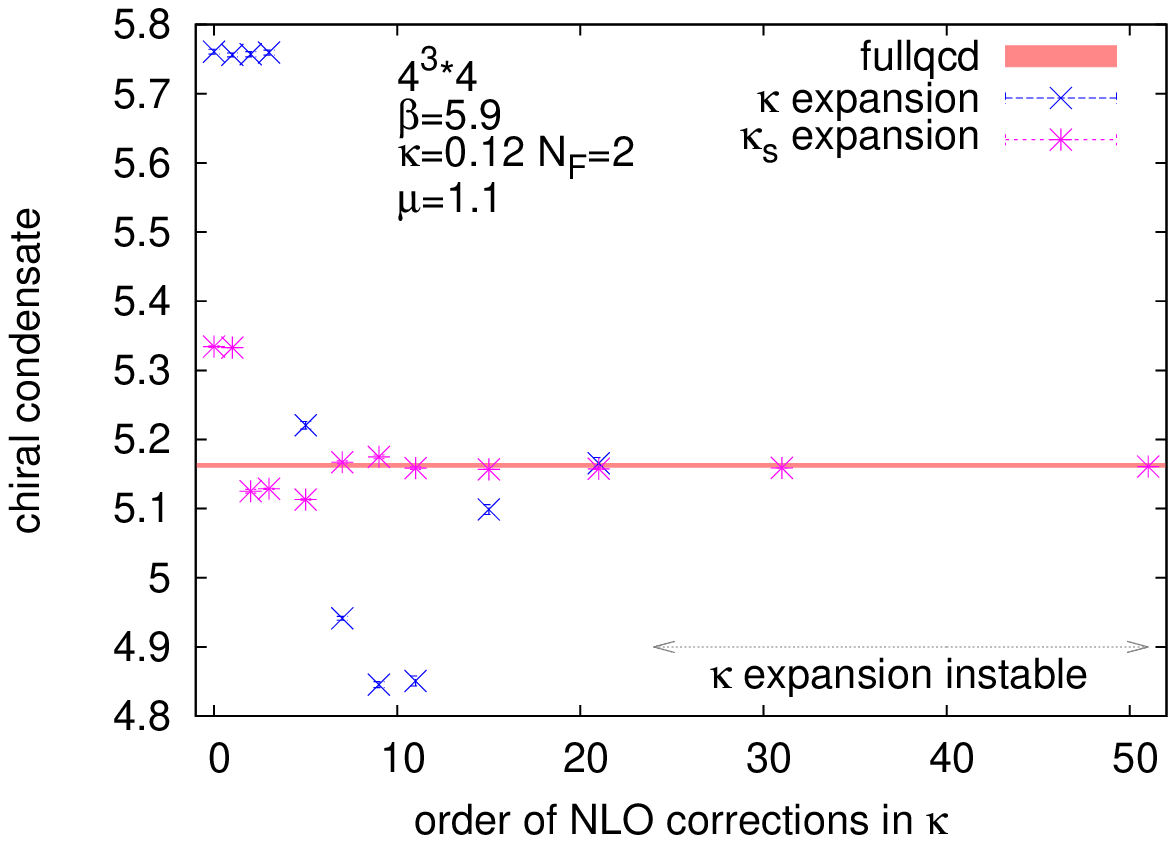}  
\includegraphics[width=0.49\columnwidth]{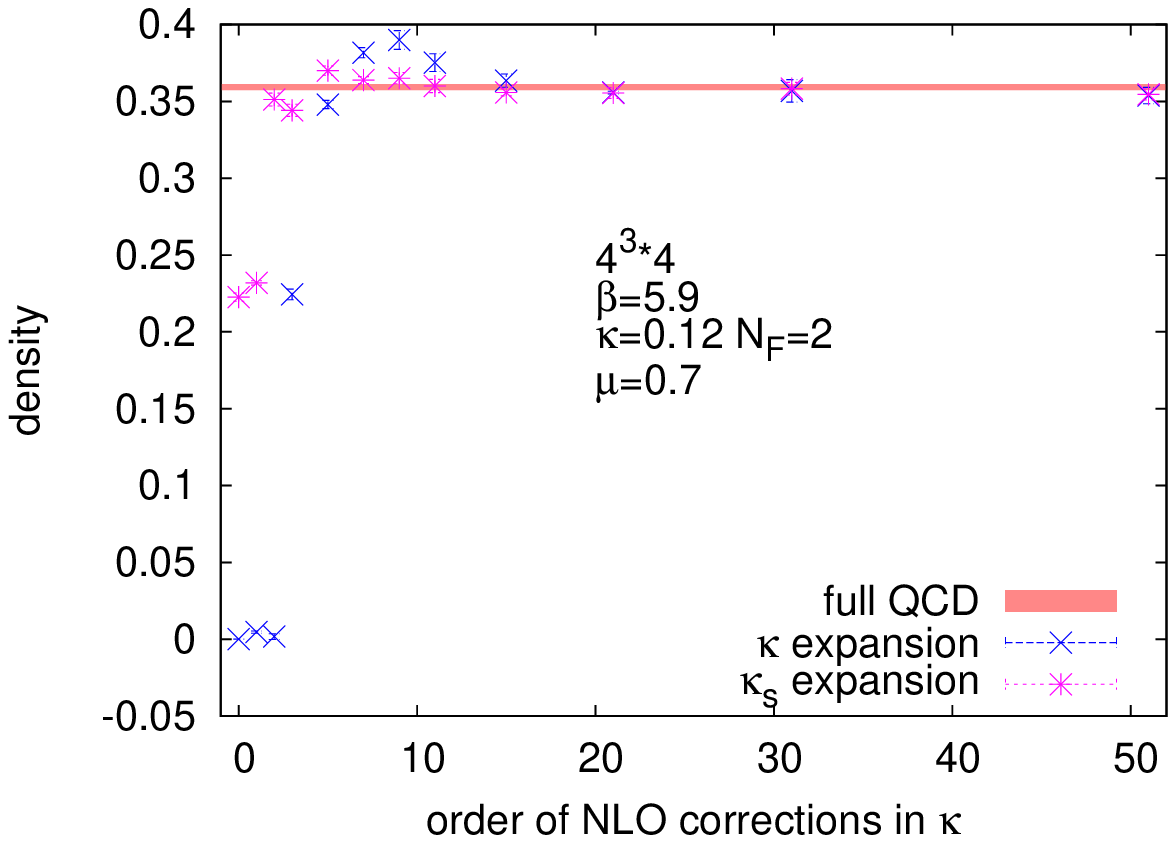}  
\includegraphics[width=0.49\columnwidth]{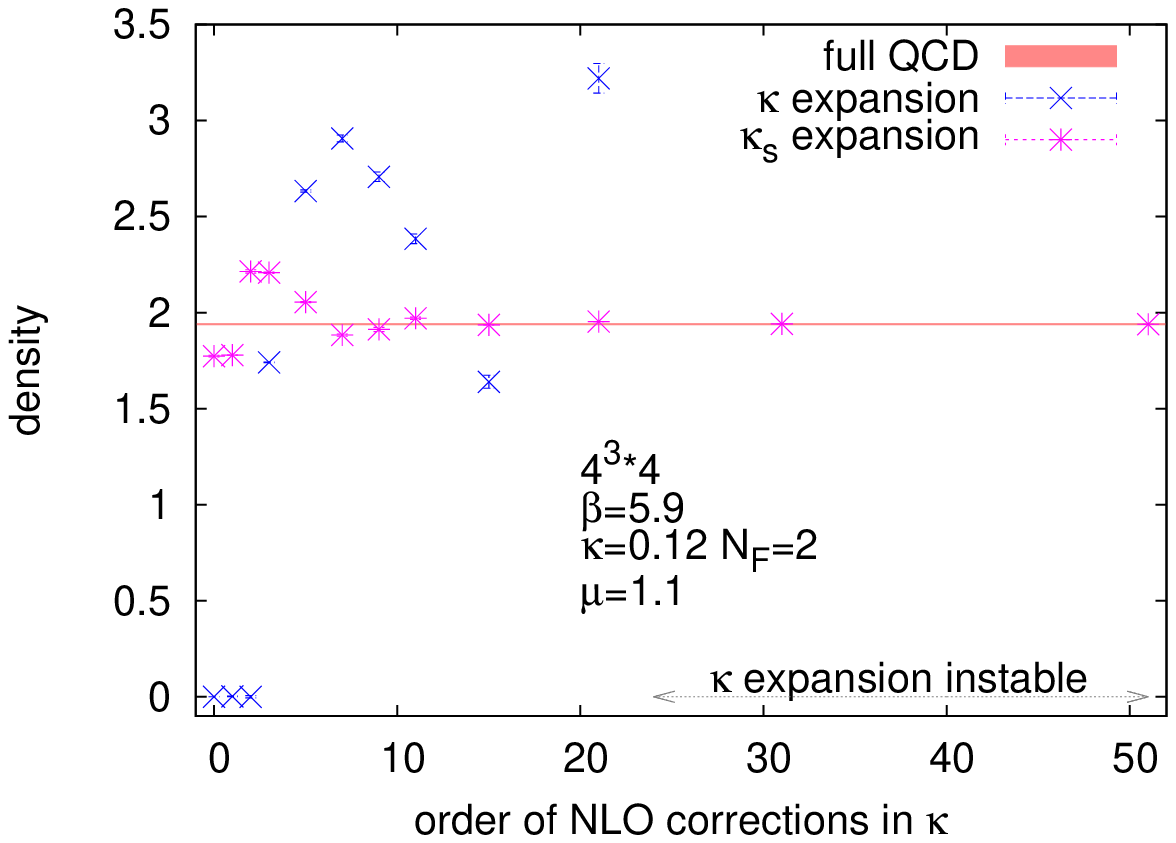}  
\includegraphics[width=0.49\columnwidth]{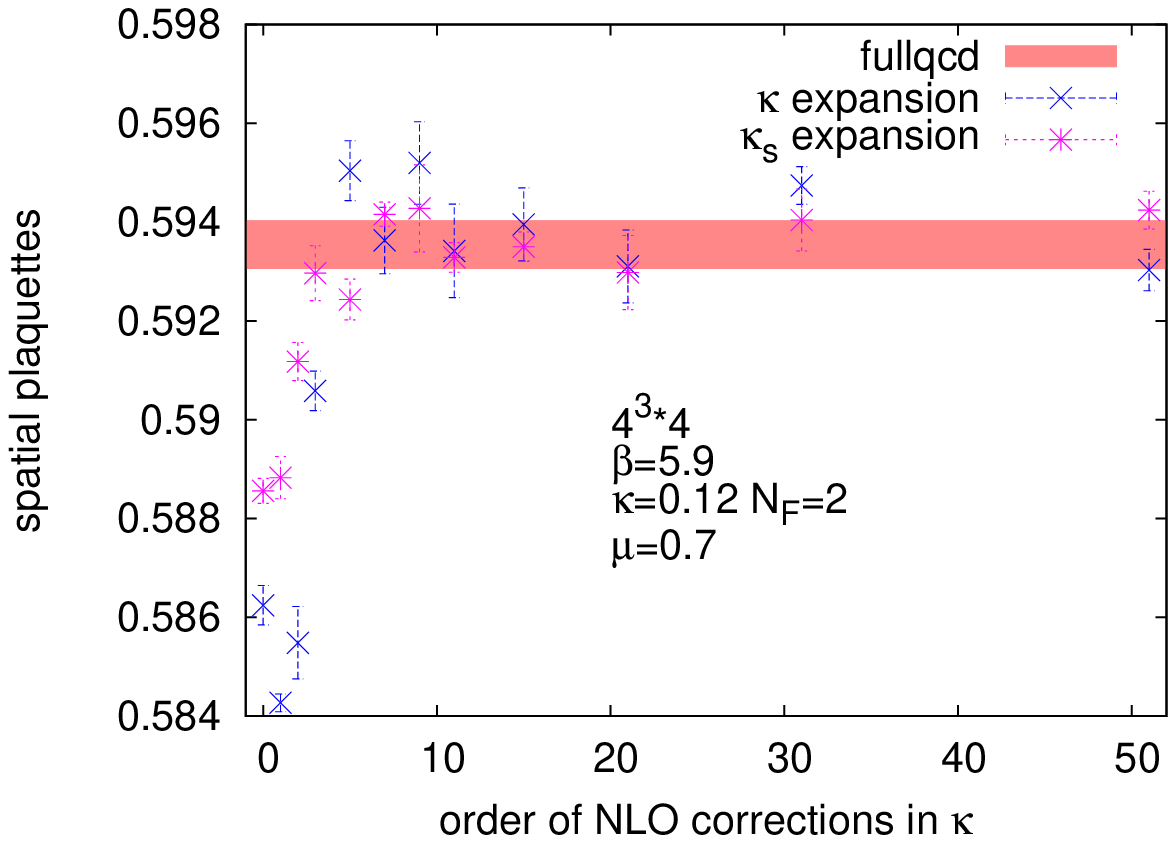}  
\includegraphics[width=0.49\columnwidth]{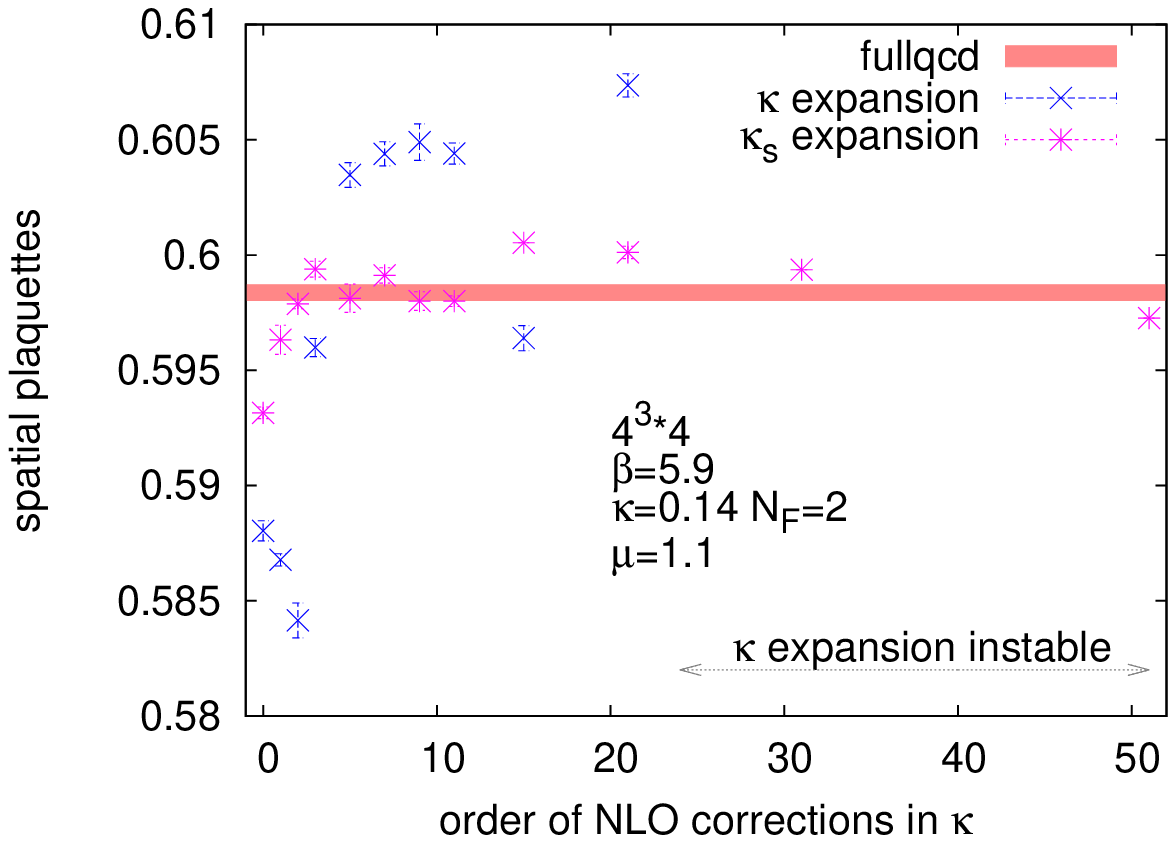}  
\caption{Convergence checks for the N$^q$LO $\kappa$- and 
$\kappa_s$-expansion on a $4^4$ lattice, $n_f=2$ 
at $\beta=5.9$, $\mu=0.7$ ({\it left}) and $\mu=1.1$ ({\it right}), vs $q$.
{\it Top}: Chiral  condensate at
$\kappa=0.12$. {\it Middle}: baryonic density at $\kappa=0.12$.
{\it Bottom}: spatial plaquette at $\kappa=0.12$ ({\it left}) and $0.14$  
({\it right}).
}
\label{f_nlopc}
\end{center}
\end{figure}

 The results 
 in Fig. \ref{f_nlopc} are obtained 
 with CLE. For  $\kappa=0.12$ the $\kappa_s$ approximation appears reasonable 
for $q \geq 8$, which is still much easier to simulate
than full QCD (while the $\kappa$-expansion is less reliable). 
For $\kappa =0.14$ some quantities need larger 
$q$ or do not converge at all. We conclude that at $\kappa=0.12$ we 
can obtain a good approximation, smaller masses need further study.

\subsection{Baryonic density at ''low" temperature.}

Here we work on lattices with $N_s \geq 10$ and $N_{\tau}  =8,10,12,16$ at
at $\beta=5.8$. Although we do not aim at physical results at this stage, we can get 
a rough
idea of the physical parameters from the scale estimate obtained by 
gradient flow
for HD-QCD in LO. This gives $l_t$ of approximately 
$1.12, 1.40, 1.68, 2.24$ fm, hence
temperatures of $\simeq \, 180, \cdots , 90$ MeV. The quarks are 
very heavy.

\begin{figure}[h]
\begin{center}
\includegraphics[width=0.49\columnwidth]{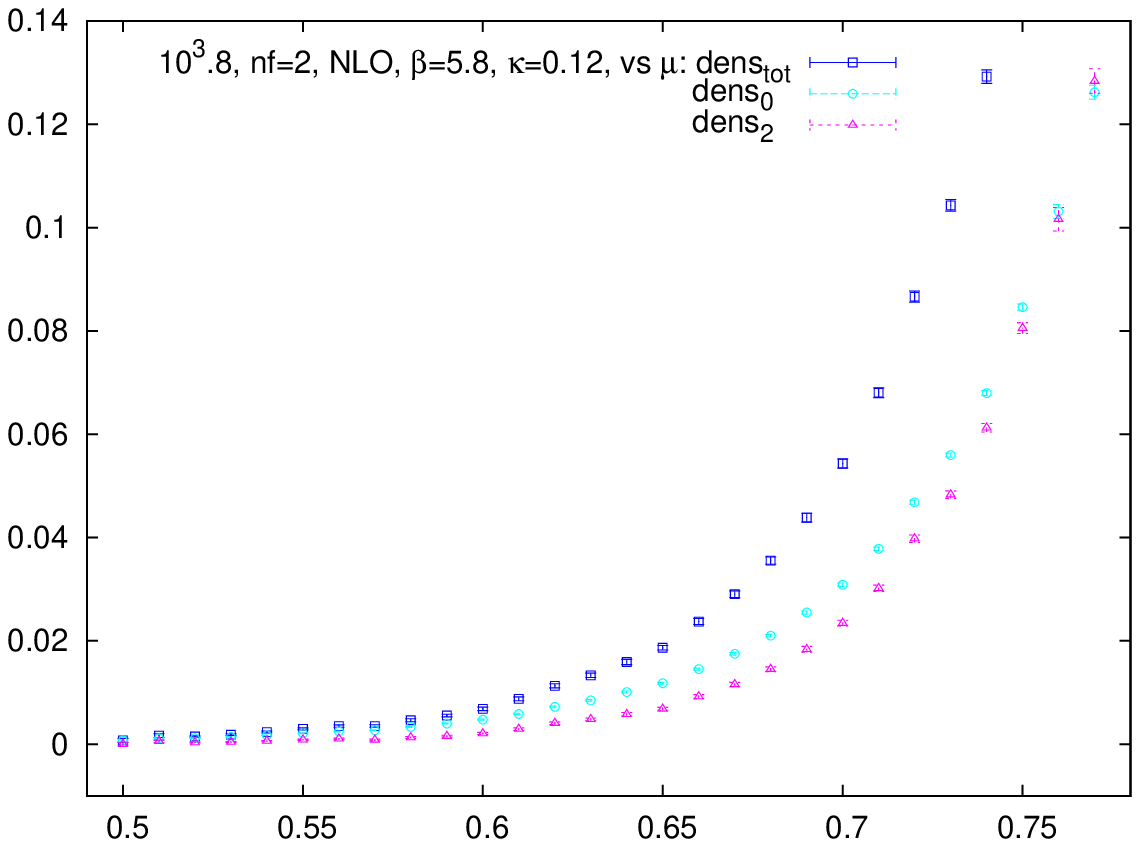}  
\includegraphics[width=0.49\columnwidth]{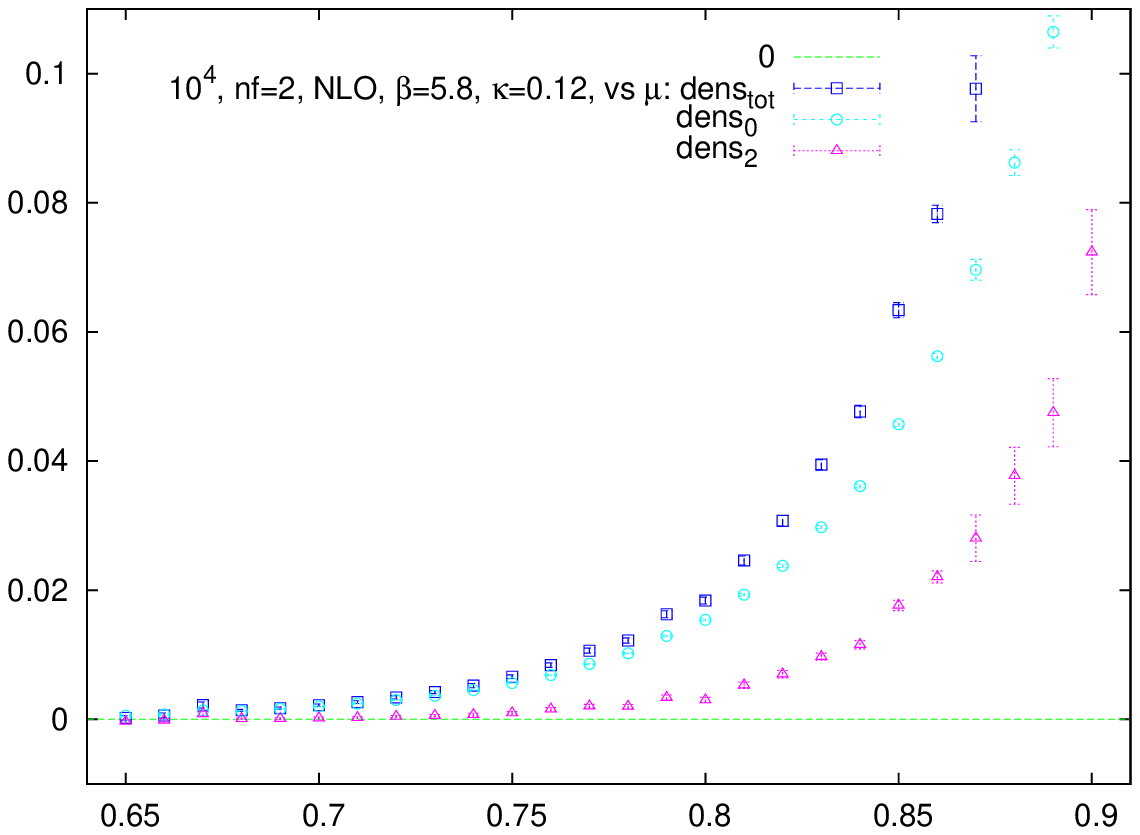}  
\includegraphics[width=0.49\columnwidth]{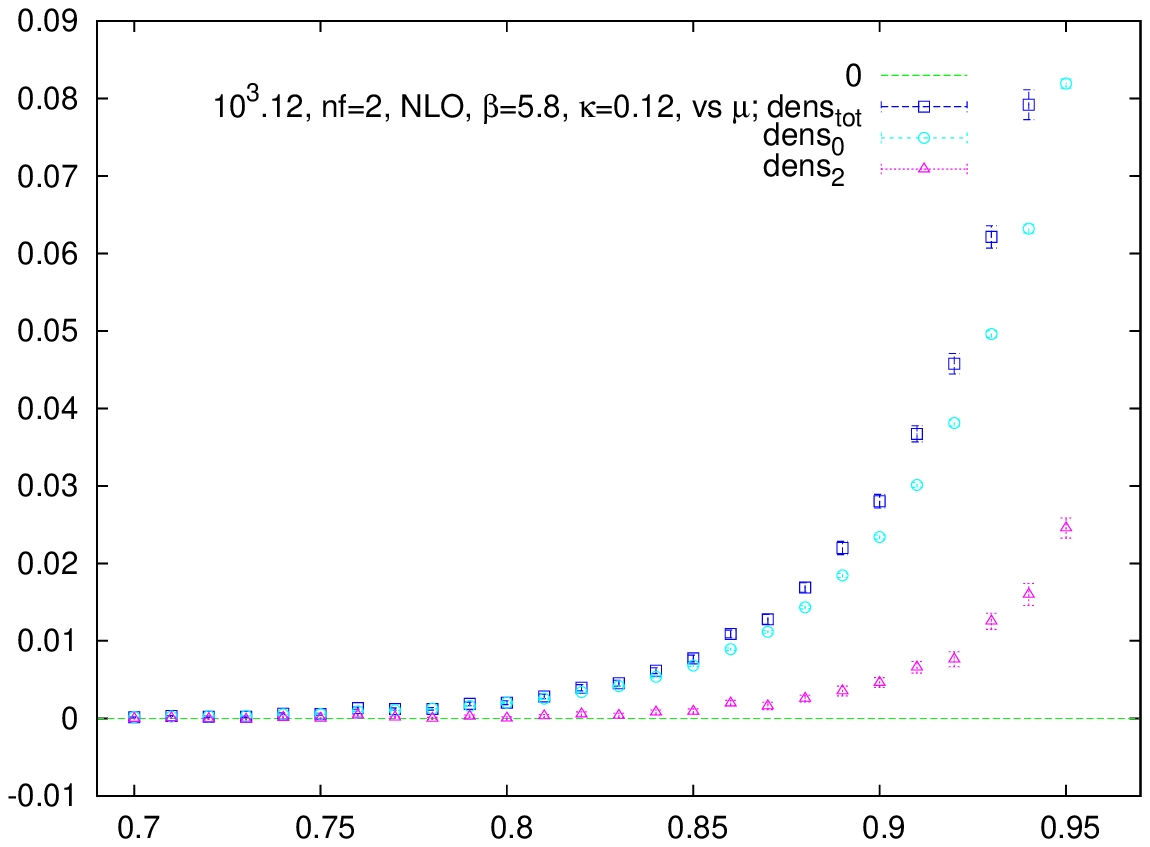}  
\includegraphics[width=0.49\columnwidth]{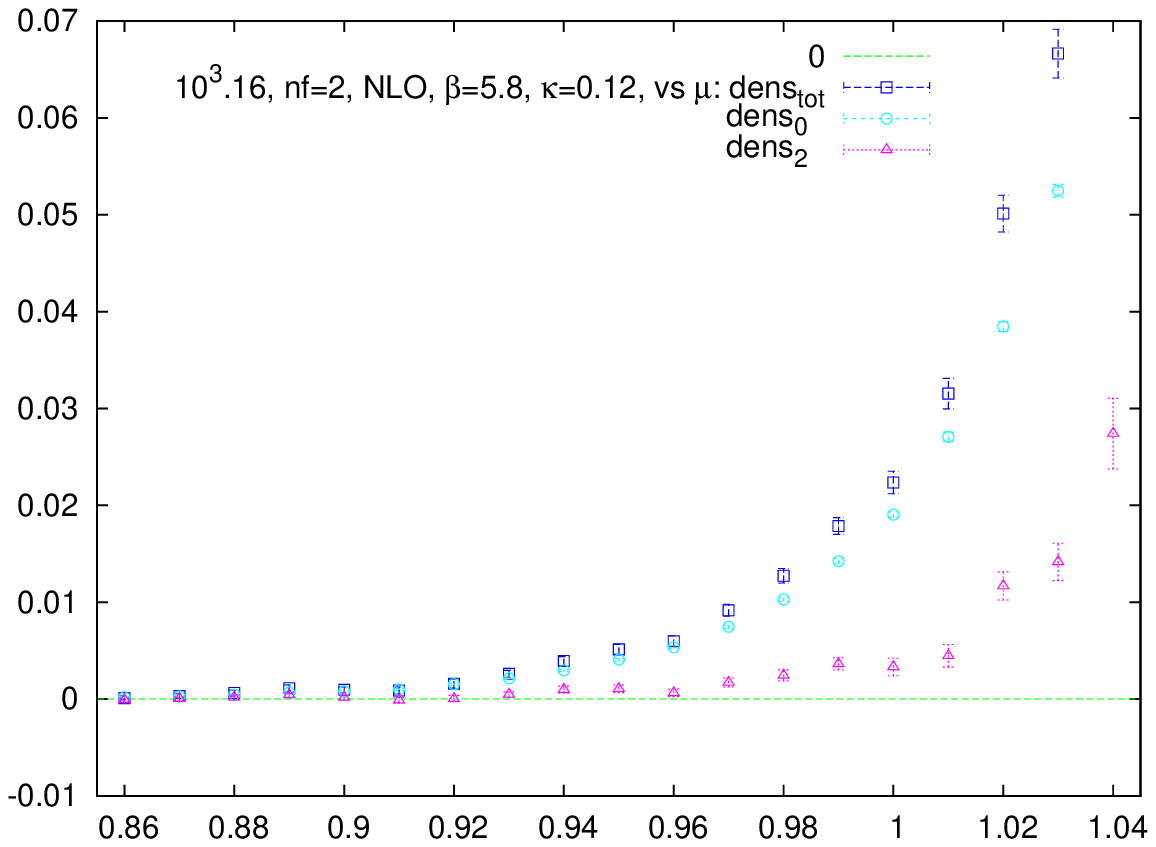}  
\caption{Baryon density from RW at $\beta =5,8$, $n_f=2$, NLO
$\kappa=0.12$ {\it Top:} 
$T \sim 180 $ and $145$ MeV.   
{\it Bottom}:
$T \sim 120 $ and $90$ MeV. Shown are also the contributions 
to the density from
the
straight and decorated Polyakov loops separately.}
\label{f_dens_8-16}
\end{center}
\end{figure}

The low $\mu$ region can be scanned by RW in first NLO. The data in 
the Fig. \ref{f_dens_8-16} are obtained with 
$N_s=10$ . Due to the small
aspect ratio we have  finite spatial size effects. Here we show only the 
baryonic density, the Polyakov loop plots have a similar behaviour. 
For the density we show separately the contributions from straight and 
decorated Polyakov loops, but notice that this is an NLO calculation,
 therefore also the former are not the same as in a LO calculation.

We observe a silver blaze region and an onset of the baryonic density 
at values of $\mu$ increasing with decreasing $T$. 
Notice that this is not the hadron-plasma
transition which, at least in LO, takes place at much larger $\mu$ 
\cite{ben14}.
Moreover we notice a seizable structure at the onset which seems to indicate 
the presence of steps. It is tempting to see here a hint for a nuclear 
matter transition at a lower temperature, not attained yet in these simulations, 
but the effects of which would propagate in the phase diagram. Since the 
temperature favours the creation of charges the dependence of the onset 
on $T$ appears realistic. 

The phase diagrams have been studied in LO in \cite{ben14} by CLE and in NLO
in \cite{dfss} by RW. They show a very similar picture, with the 
transition line curving toward smaller $\mu$ with increasing temperature, to 
become very wide below a certain $\mu$-value. This indicates that the
lower orders already catch the qualitative picture and also that 
at least in the region where they overlap both CLE and RW perform well. 
For further discussion and results see \cite{dfss} and \cite{ben14}.
For further results from CLE see \cite{asss,ben14,den14}.

\section{Summary}

We here present a program to extend the HD-QCD  to higher
orders, approaching in this way QCD within a controlable, analytic
 approximation. This 
program appears feasible, at least for not too small (bare) quark masses.
 It is based on N$^q$LO CLE simulations at $\beta > 5.7$ and for 
 various $\mu$ and lattice
 sizes and $q$ significantly larger than 1. Note also that the analytic
 properties of the expansion and of the full process are different, therefore 
 when the former  converges onto the latter we have a good test that 
 poles in the full drift are not quantitatively relevant and the results are 
 error-free.

 At $\mu$ below the hadron/plasma transition preliminary RW 
 calculations
 in NLO indicate interesting effects at the onset of the baryonic 
 density. The phase diagram has been drawn in lower orders both 
 in CLE and RW exhibiting the
 $\beta , \mu$ dependence of the hadron-plasma transition. For further 
 analysis and quantitative results we need, however, more statistics and 
 data from  CLE  which are not restricted in $\mu$ and 
 can go to larger $\kappa$ orders. \medskip

\acknowledgments
We are indebted to F. Attanasio, L. Bongiovanni and J. Pawlowski 
for discussions.
We thank the support of  BMBF and MWFK Baden-W\"urttemberg.
ES and IOS are supported by the DFG.
GA is supported by STFC, the Royal Society, 
the Wolfson Foundation and the Leverhulme Trust.

\end{document}